\documentclass[twoside,journey]{IEEEtran}
\usepackage{makecell}
\usepackage{hyperref}
\usepackage{array}
\usepackage{graphicx,amssymb,amsmath}
\usepackage{multicol}
\usepackage[noadjust]{cite}
\usepackage{setspace}
\usepackage{subfigure}
\usepackage{graphicx}
\usepackage{float}
\usepackage{url}
\usepackage{stfloats}
\usepackage{amsthm,pifont}
\usepackage{flushend}
\usepackage{cases,subeqnarray}
\usepackage{bm,multirow,bigstrut}
\usepackage{amsmath, amsthm, amssymb}
\usepackage{textcomp}
\usepackage{latexsym,bm}
\usepackage{booktabs}
\usepackage{xcolor}
\usepackage{mathtools}
\usepackage{dsfont}
\usepackage{extarrows}
\usepackage{epsfig}
\usepackage{epsfig}
\usepackage{epstopdf}
\usepackage[noend]{algpseudocode}
\usepackage{algorithmicx,algorithm}
\theoremstyle{plain}

\theoremstyle{plain}

\usepackage{amsmath}

\IEEEoverridecommandlockouts
\begin{document}

\title{Hallucination-aware Optimization for Large Language Model-empowered Communications}
\author{Yinqiu~Liu, Guangyuan Liu, Ruichen Zhang, Dusit~Niyato,~\IEEEmembership{Fellow,~IEEE},
Zehui~Xiong,\\
Dong In Kim,~\IEEEmembership{Life Fellow,~IEEE},
Kaibin Huang,~\IEEEmembership{Fellow,~IEEE}
and
Hongyang Du
\thanks{Y. Liu, G. Liu, R. Zhang, and D. Niyato are with the College of Computing and Data Science, Nanyang Technological University, Singapore (e-mails: yinqiu001@e.ntu.edu.sg, liug0022@e.ntu.edu.sg, ruichen.zhang@ntu.edu.sg, and dniyato@ntu.edu.sg).}
\thanks{Z. Xiong is with the Pillar of Information Systems Technology and Design, Singapore University of Technology and Design, Singapore (e-mail:
zehui\_xiong@sutd.edu.sg)}
\thanks{D. Kim is with the College of Information and Communication Engineering, Sungkyunkwan University, South Korea (e-mail: dongin@skku.edu).}
\thanks{K. Huang and H. Du are with the Department of Electrical and Electronic Engineering, University of Hong Kong, Hong Kong, China (e-mails: huangkb@eee.hku.hk and duhy@eee.hku.hk).}
\vspace{-0.6cm}
}
\maketitle

\begin{abstract}
Large Language Models (LLMs) have significantly advanced communications fields, such as Telecom Q\&A, mathematical modeling, and coding.
However, LLMs encounter an inherent issue known as hallucination, i.e., generating fact-conflicting or irrelevant content. 
This problem critically undermines the applicability of LLMs in communication systems yet has not been systematically explored.
Hence, this paper provides a comprehensive review of LLM applications in communications, with a particular emphasis on hallucination mitigation. 
Specifically, we analyze hallucination causes and summarize hallucination mitigation strategies from both model- and system-based perspectives.
Afterward, we review representative LLM-empowered communication schemes, detailing potential hallucination scenarios and comparing the mitigation strategies they adopted.
Finally, we present a case study of a Telecom-oriented LLM that utilizes a novel hybrid approach to enhance the hallucination-aware service experience. 
On the model side, we publish a Telecom hallucination dataset and apply direct preference optimization to fine-tune LLMs, resulting in a 20.6\% correct rate improvement. 
Moreover, we construct a mobile-edge mixture-of-experts architecture for optimal LLM expert activation.
Our research aims to propel the field of LLM-empowered communications forward by detecting and minimizing hallucination impacts.
\end{abstract}

\IEEEpeerreviewmaketitle
\vspace{-0.3cm}
\section{Introduction}
Large Language Models (LLMs), exemplified by groundbreaking developments such as OpenAI's GPT series, harness vast datasets and intricate neural networks to achieve human-like linguistic abilities \cite{wirelessLLM, zou2024telecomgpt}. 
Beyond text generation, LLMs are reshaping how machines understand and generate multimodal content, showing transformative impacts across diverse sectors, including legal, healthcare, etc. \cite{zou2024telecomgpt}.

Telecommunication researchers and companies, such as AT\&T, are paying increasing attention to LLMs. 
Given the outstanding generation capability, LLMs are widely applied to automate communication tasks such as Telecom Q\&A, coding, and network optimization. 
For instance, Zou \textit{et. al.} presented TelcomGPT, the state-of-the-art Telecom-oriented LLM for math modeling, Telecom document analysis, etc. \cite{zou2024telecomgpt}.
Shao \textit{et. al.} developed WirelessLLM, leveraging LLMs to realize wireless resource allocation, spectrum sensing, and protocol understanding \cite{wirelessLLM}.
Furthermore, Jiang \textit{et. al.} applied LLMs as the global knowledge base between senders and receivers of Semantic Communication (SemCom), using rich pre-training knowledge to improve communication efficiency \cite{10670195}.


Despite these advancements, LLMs suffer from an inherent issue called \textit{Hallucination}, which refers to generating irrelevant, self-conflicting, or fabricated content \cite{10.1145/3571730}. 
Hallucination is attributed to multiple factors, including limited reasoning abilities, insufficient or skewed communications data for training, and complex communication tasks \cite{yao2024llmlieshallucinationsbugs}. 
This issue becomes particularly problematic in LLM-empowered communications, as it leads to miscommunications, erroneous responses, and unreliable decision-making processes. For example, in Telecom Q\&A, hallucination may result in incorrect information retrieval, incorrect semantic mappings between queries and responses, or errors in protocol interpretation, directly compromising the efficiency and reliability of communication services \cite{10.1145/3571730}.
Nonetheless, existing research on LLM-empowered communications does not systematically discuss this issue, leaving a gap in the literature.

\textit{To this end, this article is the first to explore the integration of LLMs in communications with a particular focus on hallucination mitigation.} 
Specifically, we first analyze the LLM lifecycle and list the major causes of hallucinations.
We also summarize representative hallucination mitigation strategies, including both model-based approaches, such as prompt engineering and Retrieval Augmented Generation (RAG), and system-based methods, such as Mixture-of-Experts (MoE) \cite{10679152}.
Afterward, we review existing LLM-empowered communication schemes, evaluating the sources of potential hallucinations and comparing the mitigation strategies they applied. 
Noticing that existing works merely adopt model- or system-based approaches separately, we present a Telecom-oriented LLM as the case study.
Particularly, we propose a novel hybrid approach to perform hallucination-aware optimization on the user-side LLM service experience.
The contribution of this paper can be summarized as follows.
\begin{itemize}
    \item We conduct a systematic analysis of the LLM lifecycle, detailing the causes of hallucinations at each stage. Correspondingly, we summarize major strategies for mitigating hallucinations, categorized into two approaches: model-based strategies, which involve Nature Language Processing (NLP) algorithms to enhance LLMs, and system-based strategies, which focus on optimizing LLM deployment, resource allocation, security, etc.
    \item We review the existing LLM-empowered communication schemes, such as Telecom agents and LLM-based semantic communications \cite{wirelessLLM}, particularly from the viewpoint of how hallucinations may occur and which strategies that they apply for hallucination mitigation.
    \item We present a Telecom-oriented LLM with a novel hybrid approach to perform hallucination-aware optimization of the LLM service experience. First, we publish a Telecom-specific hallucination dataset based on TeleQnA and apply Direct Preference Optimization (DPO) to fine-tune LLMs, increasing the correct rate by 20.6\%. Moreover, we construct a mobile-edge MoE architecture to optimize LLM selection that jointly considers hallucination rate, resource consumption, and service latency.
\end{itemize}
\begin{figure*}[t]
\centering
\includegraphics[width=1.78\columnwidth]{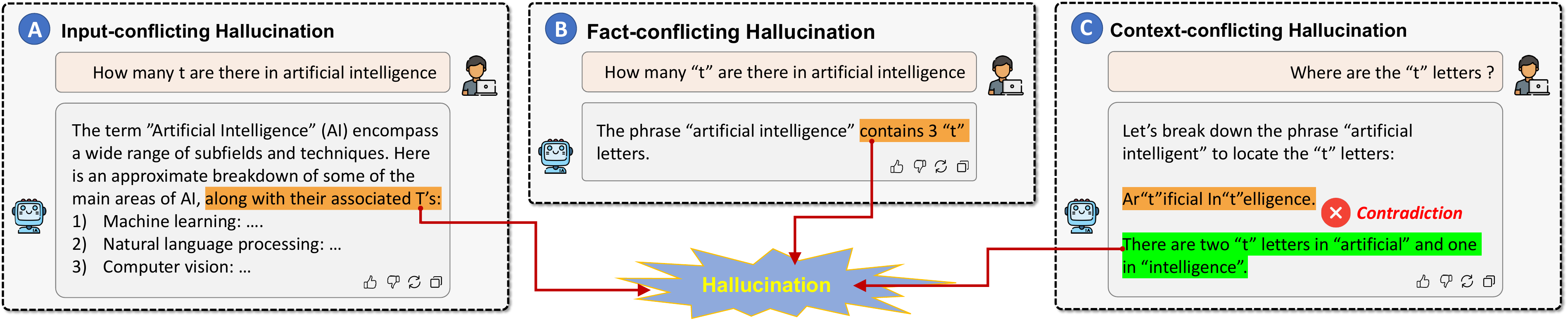}
\vspace{-0.2cm}
\caption{Practical examples of LLM hallucination. (\textbf{A}) is generated by Llama-7b Chat. (\textbf{B}) and (\textbf{C}) are generated by Claude 3.5 Sonnet.}
\vspace{-0.3cm}
\label{Hallucination}
\end{figure*}

\section{Hallucination in Large Language Models}
\subsection{Definition of Hallucination}
Hallucination refers to the phenomenon in which LLMs generate information that is factually incorrect or unfaithful to the provided input. 
Hallucinations differ from errors, which might simply be syntax or symbol mistakes, and thus become a deeper issue in the LLM's understanding and generation capability.
Various taxonomies have been proposed to categorize hallucinations. 
For instance, based on \cite{10.1145/3571730}, LLM hallucinations are classified into three types:
\begin{itemize}
    \item \textbf{Input-conflicting Hallucination}: This occurs when the LLM outputs fail to adequately address the specific input provided by a user. As shown in Fig. \ref{Hallucination}(a), the user wants to know how many ``t"s are involved in ``Artificial Intelligence". The hallucinated output elaborates on various AI sub-fields rather than locating the ``t" letters, diverging from the actual query.
    \item \textbf{Fact-conflicting Hallucination}: This arises when the content generated by LLMs conflicts with verifiable facts. As shown in Fig. \ref{Hallucination}(b), the user refines the prompt, making LLM correctly understand the meaning. However, the LLM erroneously states the number of ``t" letters as being different from the actual count.
    \item \textbf{Context-conflicting Hallucination}: This occurs when the LLM's output, though factually correct, conflicts with previously generated information by itself. As illustrated in Fig. \ref{Hallucination}(c), the LLM correctly identifies all the two ``t" letters, while the next sentence provides a contradictory conclusion.
\end{itemize}


\subsection{Causes of Hallucination}
Before analyzing the causes of hallucination, we first review the lifecycle of LLMs \cite{zhang2023sirenssongaiocean}. 
As shown in Fig. \ref{System}, the lifecycle begins with dataset collection, typically consisting of massive, often unlabeled data from open corpora (e.g., Wikipedia and Kaggle).
Afterward, LLMs undergo pre-training, where they learn to predict the next tokens based on the patterns observed in the dataset, thereby capturing a wide variety of linguistic nuances and knowledge.
The next step is Supervised Fine-tuning (SFT), which aligns the pre-trained LLMs with specific downstream tasks.
This is carried out on a smaller, often domain-specific labeled dataset or through Reinforcement Learning from Human Feedback (RLHF). 
Fine-tuning adjusts the LLM's outputs to better suit specific applications or align with user preferences, enhancing its accuracy and relevance to particular tasks.
The last stage is inference, i.e., generating responses based on user inputs. 
Here, we analyze the issues that may lead to hallucinations in these stages.
\subsubsection{Biased or Insufficient Training Data} 
LLMs trained on biased, incomplete, or insufficient datasets, e.g., collected by mobile devices, can generate hallucinations, as their responses are from skewed or limited corpora. This issue is critical for LLMs trained on the diverse and unfiltered content of the Internet, which may contain biased, incorrect, and even harmful information. In mobile contexts, this problem can be exacerbated when LLMs are trained locally on devices with user-specific data that lack verification and diversity.

\subsubsection{Model Architecture} 
Increasing the LLM size enhances its understanding and learning capability. However, when deploying on mobile devices, LLMs often undergo pruning and quantization to meet computational and storage limitations, which can impair the model's ability to generalize effectively and lead to hallucinations.
    
\subsubsection{Limited Domain Knowledge} 
LLMs operating outside their primary training domains or tasks may generate incorrect outputs due to a lack of appropriate expertise and understanding of the unseen domain. The relevance of professional knowledge may diminish over time as it becomes outdated and irrelevant, especially in fast-evolving fields that mobile users frequently access.
   
\subsubsection{Low-quality Prompts} 
Ineffective or ambiguous user inputs, as illustrated in Fig. \ref{Hallucination}(a), may fail to clearly express the intended task, leading to outputs that misinterpret or overlook the user's actual demands. This is prevalent in mobile interactions where input methods (e.g., speech or text) might constrain user expressiveness.

\subsubsection{Adversarial Attacks} 
Adversarial attacks may occur at any stage of the LLM lifecycle. For example, a backdoor attack can trigger an LLM to generate undesirable outputs by inserting poisoned samples in SFT datasets. Moreover, hallucination attacks \cite{yao2024llmlieshallucinationsbugs} demonstrate that even nonsensical, out-of-distribution prompts composed of random tokens can deceive LLMs into generating responses that appear factual but are actually false.

\begin{figure*}[t]
\centering
\includegraphics[width=1.6\columnwidth]{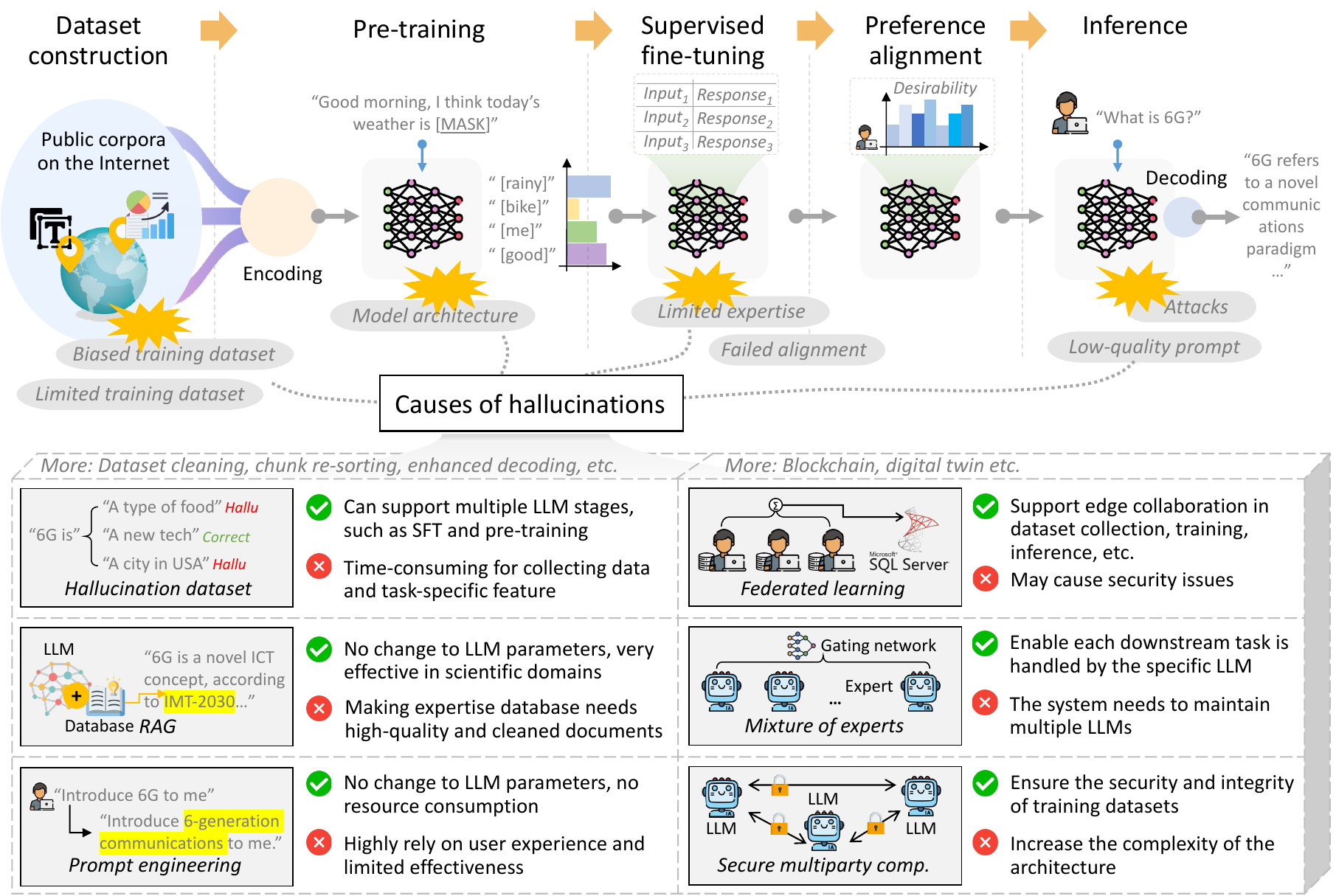}
\vspace{-0.2cm}
\caption{The illustration of LLM lifecycles and the causes of hallucinations. Also, the advantages \& disadvantages of representative model- and system-based hallucination mitigation strategies are analyzed.}
\vspace{-0.3cm}
\label{System}
\end{figure*}

\section{Hallucination Mitigation Strategies}
Hallucination can be mitigated from two perspectives.
First, by applying NLP techniques, such as RAG, fine-tuning, and prompt engineering, the performance of individual LLM can be improved.
Furthermore, in a practical LLM-empowered communication system, some architectural refinements, such as Federated Learning (FL) and MoE, contribute to reducing hallucination through efficient collaborations.
In this section, we summarize representative model- and system-based strategies to mitigate hallucinations (see Fig. \ref{System}).

\subsection{Model-based Strategies}
\subsubsection{Hallucination Detection Dataset}
Hallucination datasets \cite{MHall} gather both correct and hallucinatory outputs (manually or automatically labeled) for each input. 
By performing SFT on hallucination datasets, an LLM can be trained to increase the likelihood of correct outputs while decreasing the likelihood of hallucination. Apart from SFT, hallucination datasets can be applied in pre-training and benchmarking. \cite{MHall}.

\subsubsection{Retrieval Augmented Generation (RAG)}
RAG enhances the LLM's capacity by directly integrating external knowledge sources in real-time during the inference. By enabling the LLM to cross-verify facts and refer to up-to-date knowledge, this method significantly reduces fact-conflicting hallucinations. In particular, RAG is suitable for downstream tasks that require domain-specific professional expertise, such as legal or scientific fields.

\subsubsection{Prompt Engineering}
Utilizing structured prompts such as Chain-of-Thought (CoT) \cite{huang2023survey} can guide LLMs through a step-wise reasoning process. This method not only facilitates LLMs to adhere to a logical path but also significantly reduces the likelihood of producing contextually inappropriate or illogical outputs. By fostering a more structured analysis before generating responses, both context and input-conflicting hallucinations can be effectively mitigated. 

\subsubsection{Others}
As illustrated in Fig. \ref{System}, first, the pre-training dataset can be filtered by hash-based algorithms to reduce duplication bias. This minimizes the risk of overfitting to repeated or skewed data \cite{huang2023survey}. 
In the pre-training stage, optimizing knowledge chunk order in GPUs encourages LLMs to reason across document boundaries, improving the logical consistency among multiple generated segments. 
During the fine-tuning phase, employing LLM to assist human laborers in identifying overlooked flaws or aggregating multiple human preferences reduces sycophancy. 
Finally, factuality-enhanced decoding identifies a direction in the activation space associated with factually correct statements and then adjusts activations
along the truth-correlated direction during inference \cite{huang2023survey}.

\subsection{System-based Strategies}
\subsubsection{Federated Learning (FL)}
Applying FL allows for decentralized pre-training/SFT of LLMs across diverse datasets at various nodes in a network. 
This method not only broadens the dataset to include a wider array of examples and scenarios but also aids in minimizing data skewness and model bias, which are common sources of hallucinations. 
Moreover, FL contributes to alleviating the impact of malicious attacks by diversifying the sources of data, thus reducing the potential influence of any single maliciously manipulated data source.
\begin{table*}[htpb]
\centering
\caption{The hallucination mitigation strategies adopted by existing LLM-empowered communication schemes.} 
\includegraphics[width=0.71\textwidth]{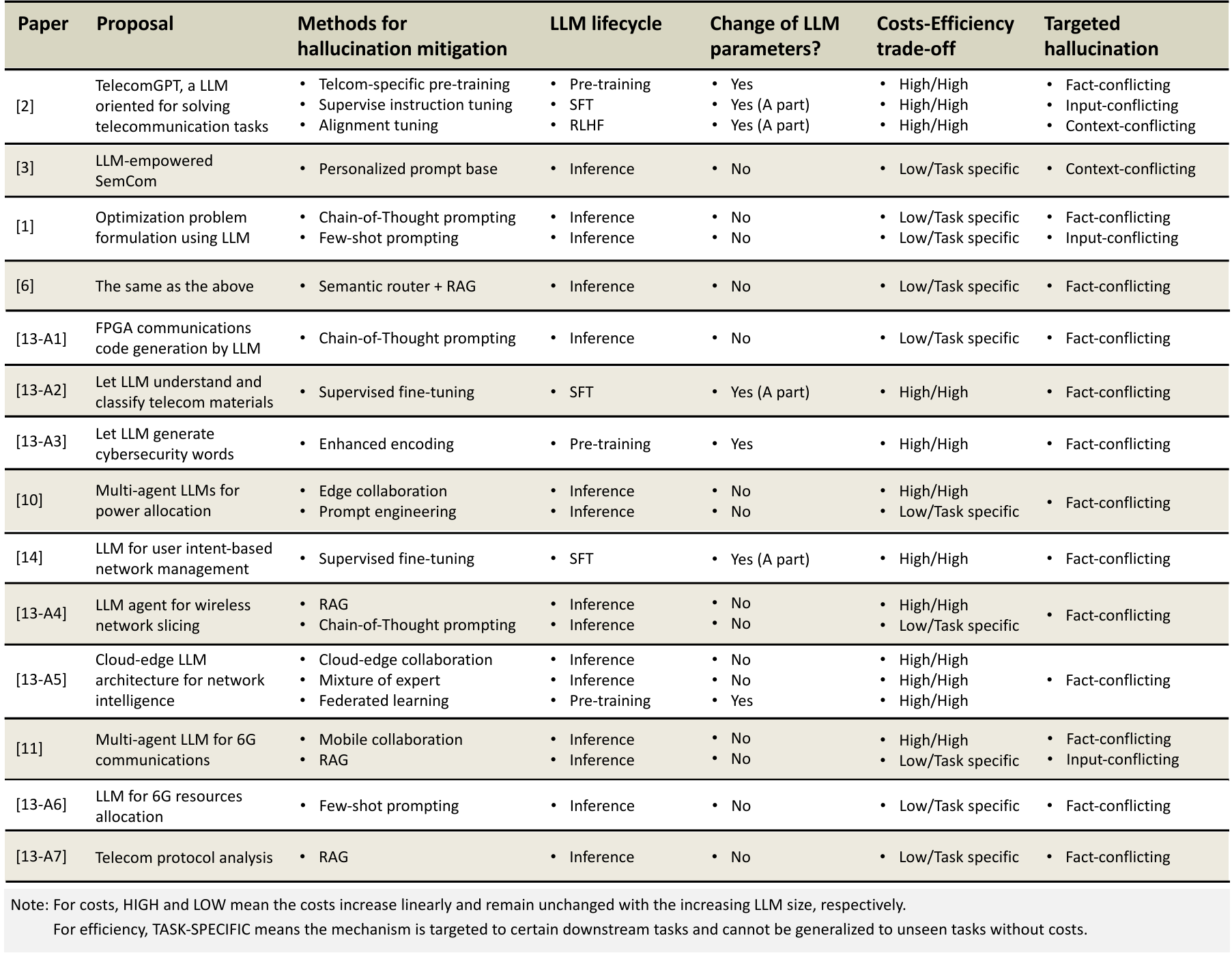}
\vspace{-0.5cm}
\label{survey}
\end{table*}

\subsubsection{Mixture of Experts (MoE)}
Employing the MoE architecture \cite{10679152} facilitates the specialization of different components of LLMs towards specific tasks or types of data. 
This specialization allows for more precise handling of queries as each ``expert" can develop a deep proficiency in a narrow area, effectively reducing the generalization errors that often lead to hallucinations. 
Additionally, the MoE architecture incorporates a gating network that dynamically selects the most suitable expert for each specific task, thereby optimizing the response accuracy and efficiency while reducing hallucination.

\subsubsection{Secure Multi-party Computation (SMPC)}
SMPC enhances the security of LLM training, which indirectly reduces the risk of hallucinations. 
By enabling nodes within a network to jointly compute a function over their inputs while keeping those inputs private, SMPC prevents any single entity from accessing the full dataset or influencing the model output maliciously. 
This is particularly important for mitigating the risk of data poisoning attacks. 



\vspace{-0.1cm}
\section{Hallucination Issues and Mitigation in Large Language Model Empowered Communications}

\subsection{Telecom-oriented Large Language Models}
LLMs are being increasingly used in the telecommunications industry, focusing primarily on telecom-oriented Q\&A, system optimization, and code generation. 
Nonetheless, hallucinations pose a significant challenge, which often stem from a lack of domain-specific knowledge, reliance on outdated information, limited reasoning abilities, and the complexity of handling multimodal communication contents.
To this end, Zou \textit{et. al.} presented TelecomPGT \cite{zou2024telecomgpt}, which combines the following hallucination mitigation strategies.
\subsubsection{Telecom-specific Pre-training}
TelecomGPT is trained on a telecom-specific dataset, including massive IEEE research papers, standards, and codes. This specialized dataset enables TelecomGPT to establish a strong foundation in the telecommunications sector, significantly reducing the likelihood of generating factually incorrect information.

\subsubsection{Supervised Instruction Tuning}
Then, TelecomGPT undergoes fine-tuning with annotated datasets tailored for specific telecom tasks, such as Q\&A and code generation. This step greatly mitigates input-conflicting hallucination by ensuring that the LLM's responses are not only contextually relevant but also closely aligned with the specific user inputs.

\subsubsection{Alignment Tuning} Since TelecomGPT tends to generate repeated or too-short responses, the authors further align model generation with annotated human preference. By quantitatively measuring the desirability of each response and guiding LLM to optimize generations, the context-conflicting hallucination can be mitigated.

\vspace{-0.1cm}
\subsection{Semantic Communications by Large Language Models}
LLMs significantly enhance SemCom by leveraging their advanced representation capabilities to design sophisticated semantic encoders/decoders \cite{10670195}. 
Moreover, LLMs act as a global knowledge base between a sender and the receiver, facilitating an interpretation of semantic information.
This is mainly due to the extensive knowledge that LLMs acquire during pre-training on massive datasets.
However, hallucinations remain a challenge in LLM-assisted SemCom, particularly when processing complicated multimodal data. 

In \cite{10670195}, the authors presented an end-to-end multimodal SemCom framework with an LLM-empowered knowledge base. 
Particularly, they incorporate a personalized prompt base tailored to GPT-4, which includes tables containing user profiles, preferences, and other personal information. 
Additionally, it maintains a repository of gold-standard prompt-response pairs, with which the LLM can be demonstrated by few-shot prompting. 
This setup enables clear instruction to LLM with enriched prompts, significantly enhancing personalization and reducing the hallucination rate by up to 30\%. 
\begin{figure*}[tpb]
\centering
\includegraphics[width=1.65\columnwidth]{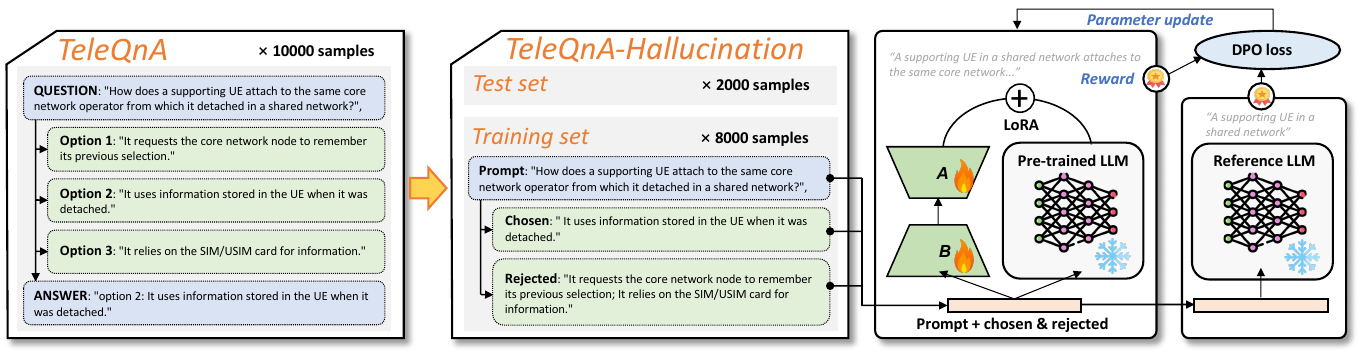}
\vspace{-0.2cm}
\caption{The data structure of TeleQnA and hallucination detection datasets and the illustration of the DPO training process. Specifically, DPO assigns \textit{chosen} answers higher rewards than \textit{rejected} ones. Then, the LoRA-enhanced LLM is trained by maximizing the gap of accumulated reward between itself and the original reference LLM.}
\vspace{-0.5cm}
\label{Dataset}
\end{figure*}

\vspace{-0.2cm}
\subsection{Large Language Model for Optimization Formulation}
Numerous communication tasks (e.g., resource allocation, service provisioning, and routing) can be modeled as an optimization problem \cite{10679152}. 
Traditionally, such formulations are performed manually, which heavily relies on human experiences. 
However, the growing heterogeneity and complexity of modern communications systems make manual formulation time-consuming, labor-intensive, and error-prone \cite{10679152}. 
Recently, LLMs have offered a promising alternative.
Specifically, their understanding capability contributes to converting human description into mathematical equations, optimization objectives, and constraints. 
Nonetheless, LLM suffers from weak reasoning capability, as they are pre-trained to predict tokens rather than derive formal formulas.
Hence, factual-, input-, and context-conflicting hallucinations may all occur.

To mitigate hallucinations, researchers leveraged prompt engineering \cite{wirelessLLM}. 
First, they demonstrated LLM-assisted power allocation modeling for the orthogonal frequency division multiplexing systems via CoT prompting.
By dividing the complex task into multiple simpler sub-tasks, the LLM's reasoning ability can be enhanced, and the generated problem formulation performs better in logic.
Additionally, they demonstrated few-shot prompting in spectrum sensing by instructing LLMs with several input-response pairs.
Such a strategy can effectively guarantee the relevance of the generation, thus reducing input-conflicting hallucinations.
Apart from prompt engineering, in \cite{10679152}, the authors adopted RAG to reduce fact-conflicting hallucinations of LLM for optimization formulation.
They created an extensive database of academic papers detailing problem-modeling processes, serving as a reservoir of domain expertise. 
Then, RAG facilitates LLM in fetching the most relevant expertise during inference, guaranteeing correct system modeling for more than 80\% cases.

\vspace{-0.2cm}
\subsection{Lessons Learned}
Apart from the aforementioned applications, LLMs are also widely adopted in network management, wireless slicing, 6G communications, etc. 
As shown in Table I, prompt engineering and RAG are performed during inferences without changing the underlying LLMs, offering cost-effective solutions. 
Prompt engineering should be utilized in human-in-the-loop communications, such as SemCom and conversational agents, where human inputs significantly determine the communication quality.
However, it cannot be relied on solely to correct deep-rooted LLM biases. 
RAG is suitable for tasks requiring significant expertise, such as optimization formulation. 
However, it is crucial to continually update the retrieval database to avoid the use of obsolete data. 
SFT is ideal when annotated domain-specific datasets are available. 
Pre-training is resource-intensive but vital for integrating hallucination mitigation early on, though the high resource demand should not be underestimated. 
Lastly, mobile-edge collaboration exploits network resources to reduce hallucination. 
Inspired by this, our case study reflects a hybrid approach that merges model- and system-based strategies to perform hallucination-aware optimization for LLM-empowered communications.

\section{Case Study: Telecom-oriented Large Language Model with Hybrid Hallucination Mitigation}

\subsection{Design Overview}
In this case study, we develop a Telecom-oriented LLM that efficiently and accurately responds to Telecom-related queries from mobile users.
Particularly, we present a novel hybrid approach to optimize our LLM.
First, DPO \cite{DPO} is adopted to fine-tune LLMs, effectively reducing fact-conflicting hallucinations by aligning outputs with predefined preferences.
Furthermore, we construct a mobile-edge MoE \cite{10679152} architecture, which organizes distributed LLM experts and activates the most appropriate experts via a Deep Reinforcement Learning (DRL)-based gating network to increase the hallucination-aware QoE of the entire communication system. 

\subsection{Hallucination Dataset Construction}
We construct a Telecom-oriented hallucination dataset atop TeleQnA \cite{github}, the first dataset specifically designed to assess the Telecom knowledge of LLMs. 
As depicted in Fig. \ref{Dataset}, TeleQnA comprises 10000 Q\&A items, each consisting of \textit{questions}, \textit{options}, \textit{answers}, and \textit{explanations}. 
In line with the prerequisites for performing DPO, we reformat TeleQnA so that each item includes one user prompt along with LLM responses, categorized into \textit{chosen} for correct responses and \textit{rejected} for hallucinatory ones (see Fig. \ref{Dataset}). 
Then, the entire dataset is randomly divided into training and testing subsets in an 80\% to 20\% ratio, respectively. 
To foster further research in detecting and mitigating hallucinations in LLM-empowered communications, we publish our dataset online \cite{github}.

\begin{figure*}[t]
\centering
\includegraphics[width=1.7\columnwidth]{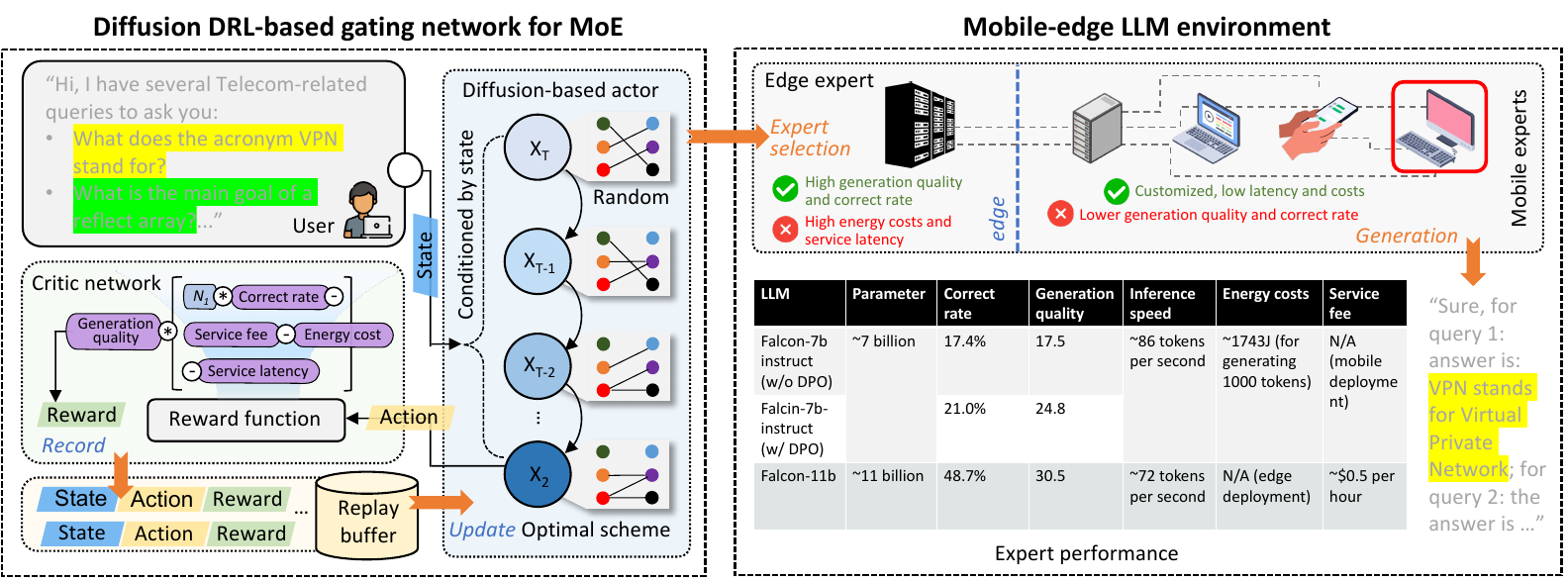}
\vspace{-0.2cm}
\caption{The diffusion DRL-based gating network (left) and the MoE architecture for maximizing hallucination-aware QoE (right). The expert performance in the table is acquired on a workstation with an NVIDIA A6000 GPU. $N_1$ in the left part refers to the number of queries in the user prompt.}
\vspace{-0.3cm}
\label{Proposal}
\end{figure*}

\subsection{Hybrid Hallucination Mitigation}
\subsubsection{Model-based Approach}
We attach Low-Rank Adaptations (LoRA) \cite{10466747} on pre-trained LLMs, which dynamically adapt LLMs by introducing trainable low-rank matrices that modify existing weights rather than retraining from scratch. 
Afterward, we adopt DPO \cite{DPO} to fine-tune LoRA-enhanced LLMs on our hallucination dataset, aiming to lower the hallucination rate in responses to Telecom-related queries. 
DPO operates through a binary classification mechanism that explicitly adjusts the probabilities associated with each response. 
It enhances the likelihood of generating \textit{chosen} responses by increasing their relative log probabilities while simultaneously reducing the likelihood of \textit{rejected} ones. 
Such a method can direct the LLM’s output to align closely with ground truth.

\subsubsection{System-based Approach}
With DPO, the hallucination rate of individual LLM can be reduced.
Nonetheless, the hallucination rate of the entire system is affected by various factors.
For instance, due to energy constraints, mobile users can hardly deploy advanced and large LLMs locally.
To access edge LLMs with lower hallucination rates, the service latency and service fee should be considered.
Moreover, different user inputs have heterogeneous sensitivity toward hallucinations.
To maximize hallucination-ware QoE of the entire communication system, we further present an MoE architecture, organizing multiple mobile-edge LLM experts.
A diffusion DRL-based gating network \cite{10.1109/TMC.2024.3356178} is applied (see Fig. \ref{Proposal}), which assigns the user request to the most suitable LLM experts.

The state space depicts the features of user inputs, including the number of Telecom queries, the expected service latency, and the service fee for consuming edge GPU resources and power. 
The action space is represented by an integer that indicates which LLM expert the user request is assigned to. 
Finally, we model hallucination-aware QoE as the reward function. 
Apart from the correct rate, the generation quality, service latency, and service fee are considered and fused (see \ref{Proposal}(left)).
The diffusion DRL \cite{10.1109/TMC.2024.3356178} follows actor-critic DRL architecture.
A diffusion model is utilized as an actor, which generates the expert assignment scheme through a Markov decision process.
Then, the critic network evaluates the desirability of the action via the reward function and guides the policy optimization (see Fig. \ref{Proposal}(left)). 
This algorithm has been proven to achieve high environment exploration capabilities, making it suitable for mobile-edge networks with heterogeneous LLMs.

\begin{figure*}[t]
\centering
\includegraphics[width=1.7\columnwidth]{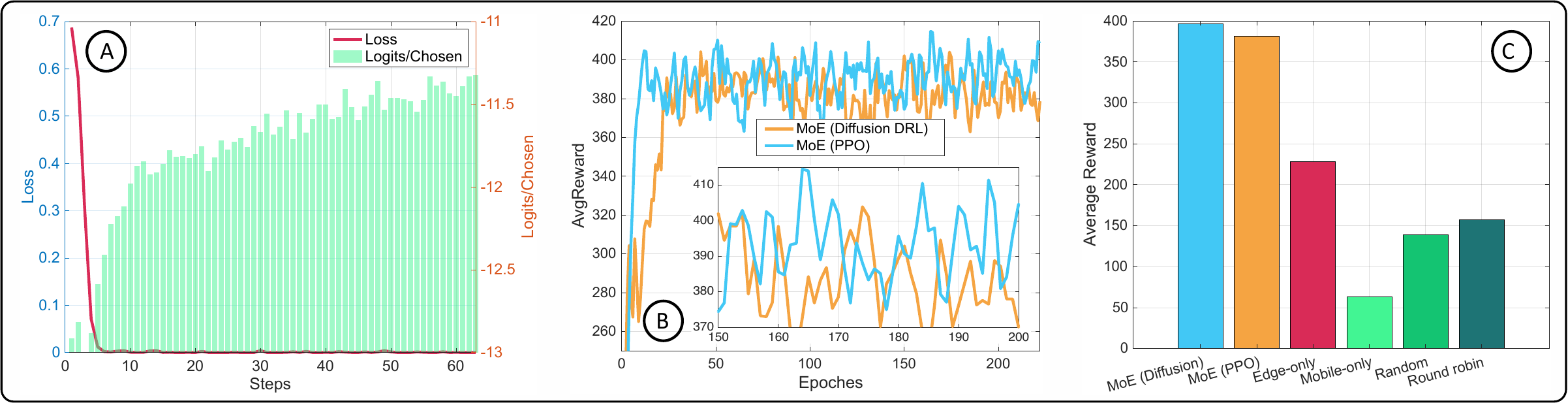}
\vspace{-0.1cm}
\caption{Experimental results. (\textbf{A}): The trends of loss and logits-chosen of DPO training. (\textbf{B}): The training curves of the MoE gating network. (\textbf{C}): The comparison of hallucination-aware QoE. Note that Edge- and Mobile-only refer to merely activating edge and mobile LLM experts, respectively. Round-robin means the experts take turns to serve users.}
\vspace{-0.4cm}
\label{Exp}
\end{figure*}

\vspace{-0.4cm}
\subsection{Experimental Results}
\subsubsection{Experimental Settings}
We employ Falcon-series LLMs to construct the mobile-edge testbed.
Specifically, mobile-level LLMs include basic Falcon-7b Instruct and Falcon-7b Instruct optimized DPO. 
The edge-level LLM is Falcon-11b, which owns higher capability due to the larger parameter size.

\subsubsection{Experimental Results}
Fig. \ref{Exp}(a) displays the DPO training curve for Falcon-7b, where the increasing logits-chosen values indicate the LLM's progressive alignment with \textit{chosen} answers. 
Afterward, we assess the correct rate (the percentage that LLM answers Telecom queries correctly without hallucination) across three LLMs.
As depicted in the table of Fig. \ref{Proposal}, DPO enhances the correct rate of Falcon-7b outputs by 20.6\%. 
Moreover, Falcon-11b demonstrates a 48.7\% correct rate due to its larger parameter size to encapsulate more complex knowledge during the pre-training phase. 
Furthermore, we illustrate the effectiveness of our MoE architecture in maximizing hallucination-aware QoE. 
Fig. \ref{Exp}(c) compares our proposal against five baselines, including mobile-only, edge-only, random, and round-robin, alongside a learning-based method named Proximal Policy Optimization (PPO). 
Among them, mobile- and edge-only approaches suffer from congestion. 
Random and round-robin strategies offer improvements by distributing loads more evenly across the network, yet they fall short of activating experts dynamically based on specific conditions and user requests. 
In contrast, our proposal significantly outperforms these methods by intelligently routing requests to the most appropriate LLM expert. 
This dynamic selection process substantially minimizes hallucinations, enhancing the efficiency of mobile-edge LLMs.

\vspace{-0.1cm}
\section{Future Directions}
\subsection{Security of LLM-empowered Communications}
LLM-empowered communication systems face security challenges at every layer. 
At the physical layer, adversarial signal injection can degrade dataset quality. 
At the network layer, routing manipulation distorts LLM inferences. 
At the application layer, data poisoning or prompt attacks lead to hallucinated outputs in tasks like Telecom Q\&A. 
The corresponding defenses require further exploration.

\vspace{-0.1cm}
\subsection{Hallucination-aware LLM Customization}
In edge networks, customization techniques such as fine-tuning, pruning, and quantization can be applied to LLMs to align with stringent resource constraints and personalized user demands. However, these might strip LLMs of their pre-trained knowledge, inadvertently increasing the likelihood of hallucinations. Therefore, it is imperative to develop hallucination-aware methods that maintain the LLM's knowledge while adapting to operational limitations.

\vspace{-0.1cm}
\subsection{LLM Reasoning}
Reasoning is a pivotal capability of LLMs, especially for tasks in communications that require logical thinking, such as math modeling and coding. Enhancing the reasoning abilities of LLMs can significantly reduce their tendency to produce hallucinatory content. Investigating methods that bolster the reasoning capabilities of LLMs (e.g., knowledge graphs) will be crucial for advancing their utility and reliability in complex communication scenarios.

\section{Conclusion}
This paper has explored LLM-empowered communications with a focus on mitigating hallucinations. 
First, we have conducted a systematic analysis of hallucination causes throughout the LLM lifecycle. 
We have then comprehensively reviewed the hallucination mitigation strategies from both model and system perspectives.
Furthermore, we have analyzed existing LLM-empowered communication schemes and compared their approaches to hallucination mitigation. 
Finally, a Telecom-oriented LLM architecture in mobile-edge networks has been presented, implementing a hybrid approach to optimize hallucination-aware QoE. 
Specifically, we have published a Telecom hallucination dataset and applied DPO to fine-tune LLMs, achieving a 20\% improvement in the correct rate. 
Additionally, a MoE architecture with a diffusion DRL-based gating network has been proposed for QoE optimization.

\bibliographystyle{IEEEtran}
\bibliography{test}

\section*{Biographies}
\indent \textbf{Yinqiu Liu} is a PhD candidate at Nanyang Technological University, Singapore.

\indent \textbf{Guangyuan Liu} is a PhD candidate at Nanyang Technological University, Singapore.

\indent \textbf{Ruichen Zhang} is a research fellow at Nanyang Technological University, Singapore.

\indent \textbf{Dusit Niyato} is a chair professor with Nanyang Technological University, Singapore.

\indent \textbf{Zehui Xiong} is an assistant professor with Singapore University of Technology and Design, Singapore.

\indent \textbf{Dong In Kim} is a professor with Sungkyunkwan University, South Korea.

\indent \textbf{Kaibin Huang} is a professor with The University of Hong Kong, Hong Kong.

\indent \textbf{Hongyang Du} is an assistant professor with The University of Hong Kong, Hong Kong.

\end{document}